\documentclass[hyper]{JHEP3}
\usepackage{graphicx,amssymb,bm,latexsym,amsmath}
\usepackage{epstopdf}
\usepackage{caption}
\usepackage{subcaption}
\usepackage[toc,page]{appendix}
\newcommand{\bej}[1]{ \begin{equation}\label{#1} }
\newcommand{\eej}{\end{equation}}
\newcommand{\beaj}[1]{\begin{eqnarray}\label{#1} }
\newcommand{\eeaj}{\end{eqnarray}}
\newcommand{\eq}[1]{(\ref{#1})}
\def\ZZZ{{\hskip-3pt\hbox{ Z\kern-1.6mm Z}}}
\def\zzz{{\hskip-3pt\hbox{ z\kern-1mm z}}}

\newcommand{\bd}{\bar{\rm D}}

\newcommand{\N}{\frac{m_{2}}{k_{2}}-\frac{m_{1}}{k_{1}}}

\newcommand{\be}{\begin{equation}}
\newcommand{\ee}{\end{equation}}
\newcommand{\ben}{\begin{eqnarray}\displaystyle}
\newcommand{\een}{\end{eqnarray}}

\def\one{{\hbox{ 1\kern-.8mm l}}}
\def\zero{{\hbox{ 0\kern-1.5mm 0}}}

\def\be{\begin{equation}}       
\def\ee{\end{equation}}         

\def\bea{\begin{eqnarray}}      
\def\eea{\end{eqnarray}}
                                
\def\ba{\begin{array}}
\def\ea{\end{array}}
\def\bd{\begin{displaymath}}
\def\ed{\end{displaymath}}

\def\eq{\begin{equation}}
\def\eqe{\end{equation}}
\def\eqa{\begin{eqnarray}}
\def\eqae{\end{eqnarray}}
\def\ena{\end{eqnarray}}

\def\unit{1 \hskip-.3em \raise2pt\hbox{$ \scriptstyle |$ } }

\def\bd{\begin{displaymath}}
\def\ed{\end{displaymath}}

\def\6{\partial}
\def\N4{{\cal N}=4}

\def\bop#1{\setbox0=\hbox{$#1M$}\mkern1.5mu
        \vbox{\hrule height0pt depth.04\ht0
        \hbox{\vrule width.04\ht0 height.9\ht0 \kern.9\ht0
        \vrule width.04\ht0}\hrule height.04\ht0}\mkern1.5mu}

\def\>{\rangle} 

\def\<{\langle} 
\def\Dsl{D \hskip-.6em \raise1pt\hbox{$ / $ } }
\def\to{\rightarrow}

\def\+{\oplus}

\def\as2{AdS_3\times S^3_1 \times S^3_2}

\title{On spinning strings in I-brane background}

\author{Sagar Biswas\\
Department of Physics, Ramakrishna Mission Vidyamandira, \\ Belur Math, Howrah 711 202, India

Email: \email{biswas.sagar09iitkgp@gmail.com}}

\abstract{We revisit semiclassical strings, in particular we focus on rigidly rotating strings, in the near horizon geometry of two orthogonal stacks of NS5-branes (I-branes) using the string sigma model. We determine the conserved charges for the probe string moving in the resulting $\mathbb{R}_t \times S^3_{\theta_1}\times S^3_{\theta_2}$ background supported by NS-NS two-forms and find a regularized dispersion relation for different values of integration constants. Using configurations that move simultaneously on both spheres, we obtain the giant magnon like dispersion relation for one particular set of parameters, while other consistent set of values gives us a dispersion relation reminiscent of the single spike.}


\begin{document}
\section{Introduction}
$AdS/CFT$ correspondence \cite{a, Gubser:2002tv, Witten:1998qj}, in its most famous avatar, relies on the duality between type IIB string states in $AdS_5\times S^5$ and certain operators in a four dimensional Conformal Field Theory (CFT) living on the boundary of $AdS_5$. The exact matching of observables from both sides of the duality has turned out to be one of the most powerful tools of modern theoretical physics. Moreover, the appearance of Integrability as a key feature in AdS/CFT correspondence makes it possible to find exact solutions of both string theory in AdS space and gauge theory on its boundary. Using the language of integrable systems, which naturally occur with elegant mathematical techniques, one can explicitly solve the spectral problem of $AdS/CFT$, i.e. the matching between string energies in the bulk with anomalous dimensions of boundary CFT operators. Integrability techniques were first developed in the SU(2) sector of planar $\mathcal{N}= 4$ SYM, as the action of the Dilatation operator in this sector can be reduced to an integrable Heisenberg spin chain \cite{Minahan:2002ve, Beisert:2003jj, Beisert:2003yb}. Followed by this, there has been a number of studies put forth to establish the appearance of integrability in other lower dimensional counterparts of AdS/CFT duality \cite{Minahan:2008hf}-\cite{purens2}. 
\medskip

In conjunction to the advent of integrability techniques in the last two decades, study of the rigidly rotating strings in the semiclassical approximation has turned out one of the interesting areas of research. In this connection the so called Hofman-Maldacena (HM) limit \cite{Hofman:2006xt}, which can be seen as a large angular momentum limit, simplifies considerably the problem of finding out the spectrum on both sides of the duality. In this particular instance, the spectrum consists of an elementary gauge theory excitation known as magnon which propagates with a conserved momentum $p$ along the associated spin chain, and this is dual to a certain rotating closed string in the $S^3 \in AdS_5$. Furthermore, a more general class of rotating string solution in $AdS_5$ is the spiky string which describes the higher twist operators from the dual field theory viewpoint \cite{Kruczenski:2004wg} and magnon solutions can be thought of as a subspace of these spike solutions. In fact it was futher determined in \cite{Ishizeki:2007we} that if one solves the most general form of the equations of motion for a rigidly rotating string on a sphere one encounters two sets of solutions in different realms of the parameter space, corresponding precisely to the giant magnon and single spike solutions. 
\medskip

To understand the AdS/CFT like dualities in more general backgrounds, it is instructive to study rigidly rotating string in the gravity side which can give us invaluable information about the corresponding gauge theory operators in the boundary side. Various other explorations in different exact string backgrounds have been undertaken by many authors, and for a non-exhaustive list one can see \cite{FirstQuantizing}-\cite{Ryang}. This intrigue forms the basis of our current investigation where we study such a novel brane background. The so-called I-brane background \cite{Green:1996dd},\cite{Itzhaki:2005tu} rises as the 1+1 dimensional intersection of two orthogonal stacks of NS5-branes, with one set of branes lying along $(x^0, x^1,\cdots, x^5)$, and other set lying along $(x^0, x^6,\cdots, x^9)$ directions. Holographically the near horizon geometry of the stacks of parallel NS5 branes accommodates the non-local Little String Theory (LST) in its worldvolume \cite{Aharony:1999ks}. When all five branes are coincident, in the S-dual picture, the near horizon geometry is given by
$$R^{2,1} \times R_{\phi} \times SU(2)_{k_1} \times SU(2)_{k_2}$$
where, $R_{\phi}$ is one combination of the radial directions away from the two sets of NS5-branes, and the coordinates of $R^{2,1}$ are $x^0, x^1$ and another combination of two radial directions. The two $SU(2)$s with levels $k_{1,2}$ describe the angular three-spheres corresponding to $(R^4)_{2345}$ and $(R^4)_{6789}$. As mentioned in \cite{Itzhaki:2005tu} this background exhibits a higher Poincare symmetry, ISO(2,1), than the expected ISO(1,1) and twice as many supercharges one might expect. Various aspects of such intersecting brane solutions in string theory has been explored in the literature \cite{Hung:2006nn}-\cite{Nayak:2010bw}.
\medskip

Semiclassical strings on I-brane have been studied by probing  the geometry with both fundamental strings and D1-strings in \cite{Kluson:2005eb}-\cite{Kluson:2007st}, and the dual single spike and giant magnon solutions have been constructed by studying rigidly rotating fundamental strings using Nambu-Goto action in \cite{Biswas:2012wu}. A similar class of semiclassical solution was obtained in its S-dual picture i.e., by studying a D1-string on two orthogonal stacks of D5-branes in \cite{Banerjee:2016avv}, which mimics the behavior of F1 strings in stacks of NS5 branes.\footnote{For related spinning string solutions from Neumann-Rosochatius like integrable systems, one can refer to \cite{Chakraborty:2022eeq}.} But, the analysis in the above two cases had a caveat, as in both the cases one could not decouple the two spheres in the near horizon limit and could only study the string dynamics under some restrictions imposed on the equations of motion. For example, in \cite{Biswas:2012wu} we could only study (i) the string rotating in one of the sphere $(\theta_1 = \theta)$ by setting the other to zero $(\theta_2=0)$ and (ii) the string rotating around both the spheres by setting $(\theta_1=\theta_2=\theta)$. 
\medskip

In case (i) we obtain the dispersion relations as,
\begin{eqnarray}
    \tilde{E} - J &=& \sqrt{K^2 - \frac{3\lambda}{\pi^2}\sin^2 \frac{\Delta\phi}{2}} \ ,  ~~~~~~~~ \text{giant magnon} \label{gm1} \\ \tilde{J} &=& \sqrt{\tilde{K}^2 - \frac{3\lambda}{\pi^2}\sin^2\Big(\frac{\pi}{2} - \theta_0\Big)} \ , ~~~ \text{single spike} \label{ss1} 
\end{eqnarray}
We can identify these dispersion relations as the giant magnon and single spike relations obtained by probing F1-string in NS5-brane backgrounds (where angular momenta were denoted as $J_1$ and $J_2$) \cite{Biswas:2011wu},
\begin{eqnarray}
    \tilde{E} - J_1 &=& \sqrt{J_2^2 - \frac{3\lambda}{\pi^2}\sin^2 \frac{p}{2}} \ ,  ~~~~~~~~ \text{giant magnon} \label{gm} \\ \tilde{J}_1 &=& \sqrt{\tilde{J}^2_2 - \frac{3\lambda}{\pi^2}\sin^2\Big(\frac{\pi}{2} - \theta_0\Big)} \ , ~~~ \text{single spike} \label{ss} 
\end{eqnarray}
where the angle deficit $\Delta\phi$ was identified with magnon momentum $p$ and $\theta_0$ is related to the initial conditions on the string dynamics. Hence, when we switch off one of the sphere (by setting $\theta_2=0$) of I-brane background we obtained the dispersion relations of NS5-brane background.
\medskip

In case (ii) the dispersion relations found were,
\begin{eqnarray}
    \tilde{E} - J &=& \sqrt{K^2 + g \frac{\lambda}{\pi^2}\sin^2 \frac{\Delta\phi}{2}} \ , ~~~~~~~~ \text{giant magnon} \label{gm2} \\ \tilde{J} &=& \sqrt{\tilde{K}^2 + h\frac{\lambda}{\pi^2}\sin^2\Big(\frac{\pi}{2} - \theta_0\Big)} \ , ~~~ \text{single spike} \label{ss2} 
\end{eqnarray}
where $h$ and $g$ were complicated combinations of different winding numbers. This indicates that these classical configurations are unable to access both $SU(2)$s distinctly in the gauge theory picture. In the current paper we will remedy the situation, by focussing on the same background but using the gauge fixed sigma model for the dynamics of the probe string. As Polyakov action for the string possess more symmetries than Nambu-Goto action for the string, this will allow us to decouple the spheres and we could study the rotating string simultaneously on both the spheres without any such restrictions.
\medskip

The rest of the paper is organized as follows. In section 2, we will review the near-horizon geometry of I-brane and find the equations of motion for the rigidly rotating strings so that those are consistent with Virasoro constraints. We will also completely classify possible configurations of the string and find the associated profiles by integrating the equations of motion. In section 3, we write down the conserved charges for the string motion, and describe how the diverging charges can be regulated. Then, using different consistent values of our parameters, we arrive at regularized dispersion relations for these charges, which can be directly compared with the dispersion relation obtained in NS5-branes.  In section 4, we summarize and comment on further investigations. 

\section{Probe strings on I-brane}\label{sec2}
As mentioned in the introduction, I-branes originate from $1+1$ dimensional intersections of fivebranes that intersect on an $\mathbb{R}^{1,1}$. In what follows, we will give a brief introduction to the target space geometry associated to this background, and discuss fundamental string probes using rigidly rotating embeddings. 

\subsection{Near-horizon geometry}
The geometry of I-brane arises when $k_1$ number of NS5-branes lying along $(0,1,\cdots,5)$ intersect $k_2$ number of NS5-branes lying along $(0,1,6,\cdots,9)$ directions in (0,1)-plane. If the branes are coincident, then the type IIB supergravity solution is given by the following metric, dilaton and three-form NS-NS fields \cite{Itzhaki:2005tu},
\begin{eqnarray}
    && ds^2= -(dx^0)^2 + (dx^1)^2 + H_1(y) \sum_{\alpha=2}^5 (dy^{\alpha})^2 + 
H_2(z) \sum_{p=6}^9 (dz^p)^2 \ , \nonumber \\  e^{2\Phi} &=& H_1(y)H_2(z) \ , ~~~ H_{\alpha\beta\gamma}=-\epsilon_{\alpha\beta\gamma\delta} \partial^{\delta} H_1(y) \ , ~~~ H_{mnp}=-\epsilon_{mnpq}\partial^qH_2(z) \ , 
\end{eqnarray}
where the two harmonic functions $H_1 = 1+\frac{k_1l_s^2}{y^2}$ and $H_2 = 1+\frac{k_2l_s^2}{z^2}$ arise because of the presence of two orthogonal spheres with $y= \sqrt{\sum_{\alpha=2}^5 (y^{\alpha})^2}$ and  $z= \sqrt{\sum_{p=6}^9 (z^{p})^2}$. In the near horizon limit ($\frac{k_1l_s^2}{y^2}>>1$ and $\frac{k_2l_s^2}{z^2}>>1$), the metric and the two form NS-NS fields are given by,
\begin{eqnarray}
    && ds^2 = -(dx^0)^2 + (dx^1)^2 + k_1l_s^2 \frac{dr_1^2}{r_1^2} + k_1l_s^2d\Omega_1^2 + k_2l_s^2 \frac{dr_2^2}{r_2^2} + k_2l_s^2d\Omega_2^2 \ , \nonumber \\ && b_{\phi_1\psi_1} = 2k_1l_s^2\sin^2\theta_1 \ , ~~~  b_{\phi_2\psi_2} = 2k_2l_s^2\sin^2\theta_2 \ ,
\end{eqnarray}
with the following form of the three spheres,
$$d\Omega_1^2 = d\theta_1^2 + \sin^2\theta_1 d\phi_1^2 + \cos^2\theta_1 d\psi_1^2 \ , ~~~ d\Omega_2^2 = d\theta_2^2 + \sin^2\theta_2 d\phi_2^2 + \cos^2\theta_2 d\psi_2^2 \ . $$
These $d\Omega_{1,2}$ are the volume elements on the sphere along $(y^2,\cdots,y^5)$ and $(z^6,\cdots,z^9)$ directions respectively. To proceed further we make the following change of variables (we choose $l_s=1, k_1=k_2=N$)
\begin{eqnarray}
    \rho_1 = \ln \frac{r_1}{\sqrt{N}} \ , ~~ \rho_2 = \ln \frac{r_2}{\sqrt{N}} \ , ~~ x^0 = \sqrt{N}t \ , ~~ x^1 = \sqrt{N}x \ . 
\end{eqnarray}
The final form of the metric and the background NS-NS two form fields are given by
\begin{eqnarray} \label{metric1}
    && ds^2 = N(-dt^2 + dx^2 + d\rho_1^2 + d\theta_1^2 + \sin^2\theta_1 d\phi_1^2 + \cos^2\theta_1 d\psi_1^2 + d\rho_2^2 + d\theta_2^2 \nonumber \\ && + \sin^2\theta_2 d\phi_2^2 + \cos^2\theta_2 d\psi_2^2) \ , ~~ b_{\phi_1\psi_1} = 2N \sin^2\theta_1 \ , ~~ b_{\phi_2\psi_2} = 2N \sin^2\theta_2 \ .
\end{eqnarray}
In the rest of the paper, we will be mainly concerned with the above background in our computations. 

\subsection{String equations of motion}
In the near horizon geometry  \eqref{metric1}, to study a fundamental string coupled to NS-NS B-field, we use the Polyakov action with a WZ term,
\begin{eqnarray}
S=-\frac{\sqrt{\lambda}}{4\pi}\int d\sigma d\tau
[\sqrt{-\gamma}\gamma^{\alpha \beta}g_{MN}\partial_{\alpha} X^M
\partial_{\beta}X^N - \epsilon^{\alpha \beta}\partial_{\alpha} X^M
\partial_{\beta}X^N b_{MN}] \ ,
\end{eqnarray}
where $\lambda$ is the 't Hooft coupling,
$\gamma^{\alpha \beta}$ is the worldsheet metric and $\epsilon^{\alpha
	\beta}$ is the antisymmetric tensor defined as $\epsilon^{\tau
	\sigma}=-\epsilon^{\sigma \tau}=1$.
Variation of the action with respect to
$X^M$ gives us the following equations of motion
\begin{eqnarray}
2\partial_{\alpha}(\eta^{\alpha \beta} \partial_{\beta}X^Ng_{KN})
&-& \eta^{\alpha \beta} \partial_{\alpha} X^M \partial_{\beta}
X^N\partial_K g_{MN} - 2\partial_{\alpha}(\epsilon^{\alpha \beta}
\partial_{\beta}X^N b_{KN}) \nonumber \\ &+& \epsilon ^{\alpha \beta}
\partial_{\alpha} X^M \partial_{\beta} X^N\partial_K b_{MN}=0 \ ,
\end{eqnarray}
and variation with respect to the metric gives the two Virasoro
constraints,
\begin{eqnarray}
g_{MN}(\partial_{\tau}X^M \partial_{\tau}X^N +
\partial_{\sigma}X^M \partial_{\sigma}X^N)&=&0 \ ,\label{v1} \\ 
g_{MN}(\partial_{\tau}X^M \partial_{\sigma}X^N)&=&0 \label{v2}\ .
\end{eqnarray}
We use the conformal gauge (i.e.
$\sqrt{-\gamma}\gamma^{\alpha \beta}=\eta^{\alpha \beta}$) with
$\eta^{\tau \tau}=-1$, $\eta^{\sigma \sigma}=1$ and $\eta^{\tau
\sigma}=\eta^{\sigma \tau}=0$) to solve the equations of motion.
For studying a generic class of rotating strings we use the ansatz,
\begin{eqnarray}
    && t = \kappa\tau  \ , ~~~ x = \mu\tau \ , ~~~ \rho_1 = m_1\tau \ , ~~~ \rho_2 = m_2\tau \ , ~~~ \theta_1=\theta_1(y) \ , ~~~ \theta_2 = \theta_2(y)  \ , \nonumber \\ && \phi_1 = \nu_1(\tau + f_1(y)) \ , ~~~ \phi_2 = \nu_2(\tau + f_2(y)) \ , ~~~  \psi_1 = \omega_1(\tau + g_1(y)) \ , ~~~ \psi_2 = \omega_2(\tau + g_2(y)) \ , \nonumber \\
\end{eqnarray}
where $\nu_1$, $\nu_2$, $\omega_1$ and $\omega_2$ are the winding numbers for $\phi_1$, $\phi_2$, $\psi_1$ and $\psi_2$ coordinates respectively. The generalised variable $y = \sigma - v\tau$ takes worldsheet coordinates from $(\sigma,\tau)$ to $(y,\tau)$.
Solving $\phi_1$ and $\psi_1$ equations  using the above set of embeddings, we get,
\begin{eqnarray}
    \frac{\partial f_1}{\partial y} &=& \frac{1}{(1-v^2)}\Big[ \frac{c_1}{\nu_1\sin^2\theta_1} -\frac{2\omega_1}{\nu_1} - v\Big] \ ,  \nonumber \\ \frac{\partial g_1}{\partial y} &=& \frac{1}{(1-v^2)}\Big[ \frac{c_2}{\omega_1\cos^2\theta_1} + \frac{2\nu_1 \sin^2\theta_1}{\omega_1 \cos^2\theta_1} - v\Big] \ .
\end{eqnarray}
Again solving $\phi_2$ and $\psi_2$ equations we get,
\begin{eqnarray}
    \frac{\partial f_2}{\partial y} &=& \frac{1}{(1-v^2)}\Big[ \frac{c_3}{\nu_2\sin^2\theta_2} -\frac{2\omega_2}{\nu_2} - v\Big] \ ,  \nonumber \\ \frac{\partial g_2}{\partial y} &=& \frac{1}{(1-v^2)}\Big[ \frac{c_4}{\omega_2\cos^2\theta_2} + \frac{2\nu_2 \sin^2\theta_2}{\omega_2 \cos^2\theta_2} - v\Big] \ ,
\end{eqnarray}
where $c_{1,2,3,4}$ are integration constants. Finally, solving for $\theta_1$ and $\theta_2$ equations we get,
\begin{eqnarray}
    \frac{\partial^2\theta_1}{\partial y^2} &=& \frac{1}{(1-v^2)^2} \Big[\frac{c_1^2}{\sin^4\theta_1} - \frac{(c_2 + 2\nu_1)^2}{\cos^4\theta_1} + 3(\nu_1^2 - \omega_1^2)\Big] \sin\theta_1\cos\theta_1 \ , \label{th1} \\ \frac{\partial^2\theta_2}{\partial y^2} &=& \frac{1}{(1-v^2)^2} \Big[\frac{c_3^2}{\sin^4\theta_2} - \frac{(c_4 + 2\nu_2)^2}{\cos^4\theta_2} + 3(\nu_2^2 - \omega_2^2)\Big] \sin\theta_2\cos\theta_2 \ . \label{th2}
\end{eqnarray}
Above equations can be integrated by first multiplying (\ref{th1}) by $2\frac{\partial\theta_{1}}{\partial y}$ and (\ref{th2}) by $2\frac{\partial\theta_{2}}{\partial y}$. The left hand side then can be written as a total derivative and each term of the whole equation can be integrated quite easily as,
\begin{eqnarray}\label{dtdy}
    \Big(\frac{\partial \theta_1}{\partial y}\Big)^2 &=& c_5 - \frac{1}{(1-v^2)^2} \Big[ \frac{c_1^2}{\sin^2\theta_1} + \frac{(c_2 + 2\nu_1)^2}{\cos^2\theta_1} - 3(\nu_1^2 - \omega_1^2)\sin^2\theta_1 \Big] \label{theq1} \ , \\  \Big(\frac{\partial \theta_2}{\partial y}\Big)^2 &=& c_6 - \frac{1}{(1-v^2)^2} \Big[ \frac{c_3^2}{\sin^2\theta_2} + \frac{(c_4 + 2\nu_2)^2}{\cos^2\theta_2} - 3(\nu_2^2 - \omega_2^2)\sin^2\theta_2 \Big] \label{theq2} \ ,
\end{eqnarray}
where $c_5$ and $c_6$ are again integration constants.
\medskip

Comparing the two Virasoro constraints, we get the following relation between various constants,
\begin{equation}
    c_1\nu_1 + c_2\omega_1 + c_3\nu_2 + c_4\omega_2 = \alpha^2v \label{cond1} \ ,
\end{equation}
where $\alpha = \sqrt{\kappa^2 - \mu^2 - m_1^2 - m_2^2}$. Also, to make sure the Virasoro constraints are explicitly consistent with the equations of motion, we require:
\begin{equation}
    c_5+c_6 = \frac{1}{(1-v^2)^2} [4\nu_1^2 + 4\nu_2^2 - \omega_1^2 - \omega_2^2 + 4(c_1\omega_1 + c_2\nu_1 + c_3\omega_2 + c_4\nu_2) + \alpha^2(1 + v^2)] \label{cond2} \ .
\end{equation}

\subsection{Solving for the string profile}
Equation (\ref{theq1}) contains several integration constants ($c_1$, $c_2$, $c_5$), to fix these integration constants we use the condition, $\frac{\partial\theta_1}{\partial y} \to 0$ as $\theta_1 \to \frac{\pi}{2}$ and obtain the following relation among the integration constants,
\begin{eqnarray}
    c_2+2\nu_1=0 \ , ~~~ c_5 = \frac{1}{(1-v^2)^2}[c_1^2 + 3(\omega_1^2 - \nu_1^2)] \label{cond3}
\end{eqnarray}
 Substituting these values in (\ref{theq1}), we obtain,
\begin{equation}\label{profile1}
    \frac{\partial\theta_1}{\partial y} = \pm \frac{\sqrt{3(\omega_1^2 - \nu_1^2)}}{(1 - v^2)} \cot\theta_1 \sqrt{\sin^2\theta_1 - \alpha_1^2} \ ,
\end{equation}
where the constant $$\alpha_1 = \frac{c_1}{\sqrt{3(\omega_1^2 - \nu_1^2)}}$$
Similarly, in order to fix the integration constants ($c_3$, $c_4$ and $c_6$) in (\ref{theq2}), we use the condition, $\frac{\partial\theta_2}{\partial y} \to 0$ as $\theta_2 \to \frac{\pi}{2}$ and obtain, 
\begin{eqnarray}
    c_4+2\nu_2=0 \ , ~~~ c_6 = \frac{1}{(1-v^2)^2}[c_3^2 + 3(\omega_2^2 - \nu_2^2)] \label{cond4}
\end{eqnarray}
 Substituting these values, we get from (\ref{theq2}),
\begin{equation}\label{profile2}
    \frac{\partial\theta_2}{\partial y} = \pm \frac{\sqrt{3(\omega_2^2 - \nu_2^2)}}{(1 - v^2)} \cot\theta_2 \sqrt{\sin^2\theta_2 - \alpha_2^2} \ ,
\end{equation}
where $$\alpha_2 = \frac{c_3}{\sqrt{3(\omega_2^2 - \nu_2^2)}}.$$
    Now integrating (\ref{profile1}) and (\ref{profile2}) we get the string profiles,
\begin{eqnarray}
    y &=& \mp \frac{(1-v^2)}{\sqrt{3(\omega_1^2 - \nu_1^2)}\sqrt{1 - \alpha_1^2}} \cosh^{-1} \Big(\frac{\sqrt{1 - \alpha_1^2}}{\cos\theta_1}\Big) \label{profile3} \ , \\  y &=& \mp \frac{(1-v^2)}{\sqrt{3(\omega_2^2 - \nu_2^2)}\sqrt{1 - \alpha_2^2}} \cosh^{-1} \Big(\frac{\sqrt{1 - \alpha_2^2}}{\cos\theta_2}\Big) \label{profile4} \ .
\end{eqnarray}
Here, we obtain two copies of string profiles arise from two completely decoupled orthogonal spheres. We also note that these string profiles are independent of each other (as (\ref{profile3}) and (\ref{profile4}) are functions of only $\theta_1$ and  $\theta_2$ respectively which are coordinates of orthogonal spheres) and can exist simultaneously (as $y=\sigma - v\tau$ can take some finite value for both the profiles at any finite worldsheet time $\tau$).
\medskip

We now move on to discuss the various classes of strings that occur here for different values of the parameters introduced. Using conditions (\ref{cond3}) and (\ref{cond4}) in (\ref{cond1}) and (\ref{cond2}) we get,
\begin{eqnarray}
    (c_1 - 2\omega_1)\nu_1 + (c_3 - 2\omega_2)\nu_2 &=& \alpha^2 v \label{cond5} \ , \\ (c_1 - 2\omega_1)^2 + (c_3 - 2\omega_2)^2 + \nu_1^2 + \nu_2^2 &=& \alpha^2 (1 + v^2) \label{cond6} \ .
\end{eqnarray}
Solving (\ref{cond5}) and (\ref{cond6}) we get,
\begin{equation}
    c_1 - 2\omega_1 = \frac{\alpha^2\nu_1v \pm \nu_2\sqrt{(\alpha^2 - \nu_1^2 - \nu_2^2)(\nu_1^2 + \nu_2^2 - \alpha^2 v^2)}}{\nu_1^2 + \nu_2^2} \label{cond7} \ .
\end{equation}
From (\ref{cond7}) we can see $c_1 - 2\omega_1$ have equal roots for $\alpha^2 = \nu_1^2 + \nu_2^2$ and $\nu_1^2 + \nu_2^2 = \alpha^2 v^2$. As we will show in the following, these two roots are responsible for two classes of dynamics.

We can easily inspect out the following cases for the parameter space spanned by the constants:

\subsubsection*{Case I: $\nu_1^2 + \nu_2^2 = \alpha^2$}
In this case, we have $c_1 - 2\omega_1 = \nu_1v$ and $c_3 - 2\omega_2 = \nu_2v$.
\subsubsection*{Case II: $\nu_1^2 + \nu_2^2 = \alpha^2v^2$}
In this case, we have $c_1 - 2\omega_1 = \frac{\nu_1}{v}$ and $c_3 - 2\omega_2 = \frac{\nu_2}{v}$.
\subsubsection*{Case III: $\alpha^2 > \nu_1^2 + \nu_2^2 > \alpha^2 v^2$}
In this case, we have real roots for $c_1 - 2\omega_1$ and $c_3-2\omega_2$. If we take 
\begin{equation}
c_1 - 2\omega_1 = \frac{\alpha^2\nu_1v + \nu_2 \sqrt{(\alpha^2 - \nu_1^2 - \nu_2^2)(\nu_1^2 + \nu_2^2 - \alpha^2 v^2)}}{\nu_1^2 + \nu_2^2}
\end{equation}
 then it leads us to 
 \begin{equation}
 c_3 - 2\omega_2 = \frac{\alpha^2\nu_2v - \nu_1 \sqrt{(\alpha^2 - \nu_1^2 - \nu_2^2)(\nu_1^2 + \nu_2^2 - \alpha^2 v^2)}}{\nu_1^2 + \nu_2^2}
 \end{equation}
Again, if we take
\begin{equation}
c_1 - 2\omega_1 = \frac{\alpha^2\nu_1v - \nu_2 \sqrt{(\alpha^2 - \nu_1^2 - \nu_2^2)(\nu_1^2 + \nu_2^2 - \alpha^2 v^2)}}{\nu_1^2 + \nu_2^2}
\end{equation}
 then we have 
 \begin{equation}
 c_3 - 2\omega_2 = \frac{\alpha^2\nu_2v + \nu_1 \sqrt{(\alpha^2 - \nu_1^2 - \nu_2^2)(\nu_1^2 + \nu_2^2 - \alpha^2 v^2)}}{\nu_1^2 + \nu_2^2}
 \end{equation}



\section{Conserved Charges and Regularized Dispersion Relations}\label{sec3}
To determine the conserved charges associated to the string motion, we start from the full form of the sigma model action in background \eqref{metric1},
\begin{eqnarray}
   S &=& -\frac{\sqrt{\lambda}}{4\pi} \int d\tau d\sigma \Big[- [(\partial_{\sigma}t)^2 - (\partial_{\tau}t)^2] +  [(\partial_{\sigma}x)^2 - (\partial_{\tau} x)^2] + [(\partial_{\sigma}\rho_1)^2 - (\partial_{\tau}\rho_1)^2] \nonumber \\ && + [(\partial_{\sigma}\theta_1)^2 - (\partial_{\tau}\theta_1)^2]  + \sin^2\theta_1 [(\partial_{\sigma}\phi_1)^2 - (\partial_{\tau}\phi_1)^2] + \cos^2\theta_1 [(\partial_{\sigma}\psi_1)^2 - (\partial_{\tau}\psi_1)^2] \nonumber \\ && + [(\partial_{\sigma}\rho_2)^2 - (\partial_{\tau}\rho_2)^2]  + [(\partial_{\sigma}\theta_2)^2 - (\partial_{\tau}\theta_2)^2] + \sin^2\theta_2 [(\partial_{\sigma}\phi_2)^2 - (\partial_{\tau}\phi_2)^2] \nonumber \\ && + \cos^2\theta_2 [(\partial_{\sigma}\psi_2)^2 - (\partial_{\tau}\psi_2)^2] + 4\sin^2\theta_1 (\partial_{\sigma}\phi_1\partial_{\tau}\psi_1 - \partial_{\tau}\phi_1\partial_{\sigma}\psi_1) \nonumber \\ && + 4\sin^2\theta_2(\partial_{\sigma}\phi_2 \partial_{\tau}\psi_2 - \partial_{\tau}\phi_2\partial_{\sigma}\psi_2)\Big] \ .
\end{eqnarray}
From this action we can easily determine the conserved energy and angular momenta using the Noether procedure, and they can be written as following,
\begin{eqnarray}
    E &=& - \int \frac{\partial\mathcal{L}}{\partial(\partial_{\tau}t)} ~dy = \frac{\sqrt{\lambda}\kappa}{2\pi} \int_{-L/2}^{L/2} ~dy = \frac{\sqrt{\lambda}\kappa L}{2\pi} \ , \nonumber \\ 	D &=&  \int \frac{\partial\mathcal{L}}{\partial(\partial_{\tau}x)} ~dy = \frac{\sqrt{\lambda}\mu}{2\pi} \int_{-L/2}^{L/2} ~dy = \frac{\sqrt{\lambda}\mu L}{2\pi} \ , \nonumber \\ P_1 &=&  \int \frac{\partial\mathcal{L}}{\partial(\partial_{\tau}\rho_1)} ~dy = \frac{\sqrt{\lambda} m_1}{2\pi} \int_{-L/2}^{L/2} ~dy = \frac{\sqrt{\lambda} m_1 L}{2\pi} \ , \nonumber \\ P_2 &=&  \int \frac{\partial\mathcal{L}}{\partial(\partial_{\tau} \rho_2)} ~dy = \frac{\sqrt{\lambda} m_2}{2\pi} \int_{-L/2}^{L/2} ~dy = \frac{\sqrt{\lambda} m_2 L}{2\pi} \ , \nonumber \\
J_{1} &=&  \int \frac{\partial\mathcal{L}}{\partial(\partial_{\tau}\phi_1)} ~dy = -\frac{\sqrt{\lambda}}{2\pi(1-v^2)} \int_{-L/2}^{L/2}[3\nu_1\sin^2\theta_1 + c_1v] ~dy \ , \nonumber \\ J_{2} &=&  \int \frac{\partial\mathcal{L}}{\partial(\partial_{\tau}\phi_2)} ~dy = -\frac{\sqrt{\lambda}}{2\pi(1-v^2)} \int_{-L/2}^{L/2}[3\nu_2\sin^2\theta_2 + c_3v] ~dy \ , \nonumber \\  K_1 &=&  \int \frac{\partial\mathcal{L}}{\partial(\partial_{\tau}\psi_1)} ~dy = \frac{\sqrt{\lambda}}{2\pi(1-v^2)} \int_{-L/2}^{L/2} [\omega_1 - 2c_1 + 2\nu_1v + 3\omega_1\sin^2\theta_1] dy \ , \nonumber \\ K_2 &=&  \int \frac{\partial\mathcal{L}}{\partial(\partial_{\tau}\psi_2)} ~dy = \frac{\sqrt{\lambda}}{2\pi(1-v^2)} \int_{-L/2}^{L/2} [\omega_2 - 2c_3 + 2\nu_2v + 3\omega_2\sin^2\theta_2] dy \ .
\end{eqnarray}
\medskip

Here $E, D, P_1$ and $P_2$ diverges as $L \to \infty$. We will use the following combination of these diverging quantities,
\begin{equation}
    \sqrt{E^2 - D^2 - P_1^2 - P_2^2} = \frac{\sqrt{\lambda}\alpha L}{2\pi} \ .
\end{equation}

 To evaluate $J_1, J_2, K_1$ and $K_2$ we need to specify the values of constants $c_{1,3}$. Also the angle deficit of the two orthogonal spheres are given by,
\begin{eqnarray}
    \Delta\phi_1 &=& \nu_1\int \frac{\partial f_1}{\partial y} ~dy = \frac{1}{1-v^2} \int_{-L/2}^{L/2} \Big[\frac{c_1}{\sin^2\theta_1} - 2\omega_1 - \nu_1v\Big] ~dy \ , \nonumber \\ \Delta\phi_2 &=& \nu_2\int \frac{\partial f_2}{\partial y} ~dy = \frac{1}{1-v^2} \int_{-L/2}^{L/2} \Big[\frac{c_3}{\sin^2\theta_2} - 2\omega_2 - \nu_2v\Big] ~dy \ .
\end{eqnarray}

In what follows, we will discuss about the fate of these charges for different regime of the parameter space as we discussed before and find the dispersion relation among them. 

\subsection{Case I:  $c_1 = 2\omega_1 + \nu_1v$ and $c_3 = 2\omega_2 + \nu_2v$}
Using the above values of constants we find,
\begin{eqnarray}
    J_1 &=& \frac{\sqrt{\lambda}}{2\pi(1-v^2)} \Big[3\nu_1\int_{-L/2}^{L/2} \cos^2\theta_1 ~dy - (3\nu_1 + 2\omega_1v + \nu_1v^2) \int_{-L/2}^{L/2}~dy\Big] \ , \nonumber \\ J_2 &=& \frac{\sqrt{\lambda}}{2\pi(1-v^2)} \Big[3\nu_2\int_{-L/2}^{L/2} \cos^2\theta_2 ~dy - (3\nu_2 + 2\omega_2v + \nu_2v^2) \int_{-L/2}^{L/2}~dy\Big] \ , \nonumber \\ K_1 &=&- \frac{3\omega_1\sqrt{\lambda}}{2\pi(1-v^2)} \int_{-L/2}^{L/2} \cos^2\theta_1 ~dy  \ , \nonumber \\  K_2 &=&- \frac{3\omega_2\sqrt{\lambda}}{2\pi(1-v^2)} \int_{-L/2}^{L/2} \cos^2\theta_2 ~dy \ .
\end{eqnarray} 
As $L \to \infty$, the charges $J_1$ and $J_2$ diverges like the combination $\sqrt{E^2 -D^2 - P_1^2 - P_2^2}$. Combining this diverging combination with $J_1$ and $J_2$, we can find finite quantities. Rescaling, $\sqrt{E^2 - D^2 - P_1^2 - P_2^2}$ as,
\begin{eqnarray}
    \tilde{E_1} &=& \frac{3\nu_1 + 2\omega_1v + \nu_1 v^2}{\alpha (1-v^2)} \sqrt{E^2 - D^2 - P_1^2 - P_2^2} \ , \nonumber \\  \tilde{E_2} &=& \frac{3\nu_2 + 2\omega_2v + \nu_2 v^2}{\alpha (1-v^2)} \sqrt{E^2 - D^2 - P_1^2 - P_2^2} \ .
\end{eqnarray} 
we find,
\begin{eqnarray}
    \tilde{E_1} + J_1 &=& \frac{\sqrt{3\lambda}\nu_1}{\pi \sqrt{\omega_1^2 - \nu_1^2}} \sqrt{1 - \alpha_1^2} \ , \nonumber \\ \tilde{E_2} + J_2 &=& \frac{\sqrt{3\lambda}\nu_2}{\pi \sqrt{\omega_2^2 - \nu_2^2}} \sqrt{1 - \alpha_2^2} \ .
\end{eqnarray}
On the other hand the charges $K_1$ and $K_2$ are finite,
\begin{eqnarray}
    K_1 &=& -\frac{\sqrt{3\lambda}\omega_1}{\pi \sqrt{\omega_1^2 - \nu_1^2}} \sqrt{1 - \alpha_1^2} \ , \nonumber \\ K_2 &=& -\frac{\sqrt{3\lambda} \omega_2}{\pi \sqrt{\omega_2^2 - \nu_2^2}} \sqrt{1 - \alpha_2^2} \ ,
\end{eqnarray}
and the angle deficits are also finite,
\begin{eqnarray}
    \Delta \phi_1 = 2\cos^{-1}(\alpha_1) \ , ~~~  \Delta \phi_2 =
 2\cos^{-1}(\alpha_2) \ .
\end{eqnarray}
which implies 
\begin{eqnarray}
   \alpha_1 = \cos \frac{\Delta\phi_1}{2} \ , ~~~  \alpha_2 = \cos \frac{\Delta\phi_2}{2} \ .
\end{eqnarray}
Now these conserved charges hold the relationship 
\begin{eqnarray}
    \tilde{E_1} + J_1 &=& \sqrt{K_1^2 - \frac{3\lambda}{\pi^2} \sin^2 \Big( \frac{\Delta\phi_1}{2}\Big)} \ , \nonumber \\ \tilde{E_2} + J_2 &=& \sqrt{K_2^2 - \frac{3\lambda}{\pi^2} \sin^2 \Big( \frac{\Delta\phi_2}{2}\Big)} \ . \label{gm3}
\end{eqnarray}
If we identify angle deficits $\Delta\phi_1$ and $\Delta\phi_2$ with magnon momenta $p_1$ and $p_2$ respectively then the above relations can be identified with the giant magnon relations.  Here, we obtain two copies of giant magnon relations with $J_{1,2}$ replaced by $-J_{1,2}$ and each of this relation is of the form (\ref{gm1}) rather than (\ref{gm2}). As mentioned in the Section 1, (\ref{gm1}) was obtained by switching off one of the spheres (by setting $\theta_1=\theta$ and $\theta_2=0$), which effectively reduces the I-brane background into NS5-brane background. By comparing the giant magnon relation of \cite{Biswas:2011wu} we confirm that each of these giant magnon relations arises from the NS5-branes. Thus, we can say as the Polyakov action for string have completely decoupled the orthogonal spheres of I-brane, here we obtain two sets of giant magnon relations arising from two orthogonal stacks of NS5-branes. This type of behaviour was not seen while using the Nambu-Goto action for string on the same background under the restriction $\theta_1=\theta_2=\theta$, there instead of two sets of giant magnon relations we obtained one set of giant magnon relation of the form (\ref{gm2}) with a complicated combination of winding numbers encoded in $g$. 


\subsection{Case II: $c_1 = 2\omega_1 + \frac{\nu_1}{v}$ and $c_3 = 2\omega_2 + \frac{\nu_2}{v}$}
For $c_1 = 2\omega_1 + \frac{\nu_1}{v}$ and $c_3=2\omega_2 + \frac{\nu_2}{v}$, angular momenta $J_1, J_2, K_1, K_2$ becomes,
\begin{eqnarray}
    J_1 &=& \frac{\sqrt{\lambda}}{2\pi(1-v^2)} \Big[3\nu_1\int_{-L/2}^{L/2} \cos^2\theta_1 ~dy - 2(2\nu_1 + \omega_1v) \int_{-L/2}^{L/2}~dy\Big] \ , \nonumber \\ J_2 &=& \frac{\sqrt{\lambda}}{2\pi(1-v^2)} \Big[3\nu_2\int_{-L/2}^{L/2} \cos^2\theta_2 ~dy - 2(2\nu_2 + \omega_2v) \int_{-L/2}^{L/2}~dy\Big] \ , \nonumber \\ K_1 &=&- \frac{\sqrt{\lambda}}{2\pi(1-v^2)} \Big[3\omega_1\int_{-L/2}^{L/2} \cos^2\theta_1 ~dy + \frac{2\nu_1}{v}(1-v^2) \int_{-L/2}^{L/2}~dy\Big]  \ , \nonumber \\  K_2 &=&- \frac{\sqrt{\lambda}}{2\pi(1-v^2)} \Big[3\omega_2 \int_{-L/2}^{L/2} \cos^2\theta_2 ~dy + \frac{2\nu_2}{v}(1-v^2) \int_{-L/2}^{L/2} ~dy \Big] \ .
\end{eqnarray} 
As $L \to \infty$, all the above quantities diverge. Here also we can obtain a finite piece (regularize) by combining these diverging charges with the diverging combination $\sqrt{E^2 - D^2 - P_1^2 - P_2^2}$ as, 
\begin{eqnarray}
    (J_1)_{reg} &=& J_1 + \frac{2(2\nu_1 + \omega_1v)}{\alpha(1-v^2)} \sqrt{E^2 - D^2 - P_1^2 - P_2^2} = \frac{\sqrt{3\lambda}\nu_1\sqrt{1 - \alpha_1^2}}{\pi\sqrt{\omega_1^2 - \nu_1^2}} \ , \nonumber \\ (J_2)_{reg} &=& J_2 + \frac{2(2\nu_2 + \omega_2v)}{\alpha(1-v^2)} \sqrt{E^2 - D^2 - P_1^2 - P_2^2} = \frac{\sqrt{3\lambda}\nu_2\sqrt{1 - \alpha_2^2}}{\pi\sqrt{\omega_2^2 - \nu_2^2}} \ , \nonumber \\ (K_1)_{reg} &=& K_1 + \frac{2\nu_1}{\alpha v} \sqrt{E^2 - D^2 - P_1^2 - P_2^2} = -\frac{\sqrt{3\lambda}\omega_1\sqrt{1 - \alpha_1^2}}{\pi\sqrt{\omega_1^2 - \nu_1^2}} \ , \nonumber \\ (K_2)_{reg} &=& K_2 + \frac{2\nu_2}{\alpha v} \sqrt{E^2 - D^2 - P_1^2 - P_2^2} = - \frac{\sqrt{3\lambda}\omega_2\sqrt{1 - \alpha_2^2}}{\pi\sqrt{\omega_2^2 - \nu_2^2}} \ . 
\end{eqnarray}
In this case the angle deficits are also diverging, again subtracting the diverging piece, we obtain, 
\begin{eqnarray}
    (\Delta \phi_1)_{reg} &=& \Delta\phi_1 - \frac{2\pi\nu_1}{\sqrt{\lambda} \alpha v} \sqrt{E^2 - D^2 - P_1^2 - P_2^2} = 2\cos^{-1}(\alpha_1) \ , \nonumber \\  (\Delta \phi_2)_{reg} &=& \Delta\phi_2 - \frac{2\pi\nu_2}{\sqrt{\lambda} \alpha v} \sqrt{E^2 - D^2 - P_1^2 - P_2^2} =
 2\cos^{-1}(\alpha_2) \ ,
\end{eqnarray}
which implies 
\begin{eqnarray}
   \alpha_1 = \cos \frac{(\Delta\phi_1)_{reg}}{2} \ , ~~~  \alpha_2 = \cos \frac{(\Delta\phi_2)_{reg}}{2} \ .
\end{eqnarray}
Now these conserved charges hold the relationship 
\begin{eqnarray}
    (J_1)_{reg} &=& \sqrt{(K_1)_{reg}^2 - \frac{3\lambda}{\pi^2} \sin^2 \Big( \frac{(\Delta\phi_1)_{reg}}{2}\Big)} \ , \nonumber \\ (J_2)_{reg} &=& \sqrt{(K_2)_{reg}^2 - \frac{3\lambda}{\pi^2} \sin^2 \Big( \frac{(\Delta\phi_2)_{reg}}{2}\Big)} \ .
\end{eqnarray}
Here, we obtain two copies of single-spike relations and each of these single spike relations is of the form (\ref{ss1}) instead of (\ref{ss2}). As mentioned earlier, (\ref{ss1}) was obtained by effectively switching off one of the NS5-branes in I-brane background and exactly similar kind of relation was obtained by studying F1-strings in NS5-brane \cite{Biswas:2011wu}. With the same spirit as in the previous section, we can say as the Polyakov action for the string have completely decoupled the orthogonal spheres of I-brane, here we obtain two sets of single spike relations arising from two orthogonal spheres. This type behaviour was not seen while using the Nambu-Goto action for the string on the same background under the restriction $\theta_1=\theta_2=\theta$, there instead of two sets of single spike relations we obtained one set of single spike relation of the form (\ref{ss2}) with a complicated combination of winding numbers encoded in $h$. Two sets of single-spike relations were also obtained while studying rigidly rotating strings in curved NS5-branes \cite{Biswas:2013ela} which arises from the near horizon-geometry of NS1-NS1'-NS5-NS5' brane \cite{Papadopoulos:1999tw}.

\subsection{Case III: $c_1 = 2\omega_1 + \frac{\alpha^2\nu_1v \pm \nu_2\beta_1\beta_2}{\nu_1^2 + \nu_2^2}$ and $c_3 = 2\omega_2 + \frac{\alpha^2\nu_2v \mp \nu_1\beta_1\beta_2}{\nu_1^2 + \nu_2^2}$}
Let $\beta_1^2 = \alpha^2 - \nu_1^2 - \nu_2^2$ and $\beta_2^2 = \nu_1^2 + \nu_2^2 - \alpha^2 v^2$. Unlike previous two cases where we use $\beta_1^2=0$ and $\beta_2^2=0$, we now consider $\beta_1^2$ and $\beta_2^2$ as some positive quantities, i.e., $\alpha^2 > \nu_1^2 + \nu_2^2$ and $\nu_1^2 + \nu_2^2 > \alpha^2 v^2$, and combining them we get,
\begin{equation}
    \alpha^2 > \nu_1^2 + \nu_2^2 > \alpha^2 v^2 \ ,
\end{equation}
which implies $v^2<1$. One can easily check in this case also all the conserved charges diverges. However, one can find the finite quantities by subtracting the diverging piece as, 
\begin{eqnarray}
    (J_1)_{reg} &=& J_1 + \frac{1}{\alpha(1-v^2)} \Big[ 3\nu_1 + 2\omega_1 v + \frac{\alpha^2\nu_1v^2 \pm \nu_2 v \beta_1\beta_2}{\nu_1^2+\nu_2^2}\Big] \sqrt{E^2 - D^2 - P_1^2 - P_2^2} \nonumber \\ && = \frac{\sqrt{3\lambda}\nu_1\sqrt{1 - \alpha_1^2}}{\pi\sqrt{\omega_1^2 - \nu_1^2}} \ , \nonumber \\ (J_2)_{reg} &=& J_2 + \frac{1}{\alpha(1-v^2)} \Big[ 3\nu_2 + 2\omega_2 v + \frac{\alpha^2\nu_2v^2 \mp \nu_1 v \beta_1\beta_2}{\nu_1^2 + \nu_2^2}\Big] \sqrt{E^2 - D^2 - P_1^2 - P_2^2} \nonumber \\ && = \frac{\sqrt{3\lambda}\nu_2\sqrt{1 - \alpha_2^2}}{\pi\sqrt{\omega_2^2 - \nu_2^2}} \ , \nonumber \\ (K_1)_{reg} &=& K_1 - \frac{2}{\alpha (1 - v^2)} \Big[ \nu_1v - \frac{\alpha^2\nu_1v \pm \nu_2 \beta_1\beta_2}{\nu_1^2+\nu_2^2}\Big] \sqrt{E^2 - D^2 - P_1^2 - P_2^2} \nonumber \\ && = -\frac{\sqrt{3\lambda}\omega_1\sqrt{1 - \alpha_1^2}}{\pi\sqrt{\omega_1^2 - \nu_1^2}} \ , \nonumber \\ (K_2)_{reg} &=& K_2 - \frac{2}{\alpha (1 - v^2)} \Big[ \nu_2 v - \frac{\alpha^2\nu_2v \mp \nu_1 \beta_1\beta_2}{\nu_1^2+\nu_2^2}\Big] \sqrt{E^2 - D^2 - P_1^2 - P_2^2} \nonumber \\ && = - \frac{\sqrt{3\lambda}\omega_2\sqrt{1 - \alpha_2^2}}{\pi\sqrt{\omega_2^2 - \nu_2^2}} \ . 
\end{eqnarray}
In this case the angle deficits are also diverging, again subtracting the diverging piece, we obtain, 
\begin{eqnarray}
    (\Delta \phi_1)_{reg} &=& \Delta\phi_1 + \frac{2\pi}{\sqrt{\lambda} \alpha (1 - v^2)} \Big[ \nu_1 v - \frac{\alpha^2\nu_1v \pm \nu_2  \beta_1\beta_2}{\nu_1^2+\nu_2^2}\Big] \sqrt{E^2 - D^2 - P_1^2 - P_2^2} \ , \nonumber \\  (\Delta \phi_2)_{reg} &=& \Delta\phi_2 + \frac{2\pi}{\sqrt{\lambda} \alpha (1 - v^2)} \Big[ \nu_2v - \frac{\alpha^2\nu_2v \mp \nu_1 \beta_1\beta_2}{\nu_1^2+\nu_2^2}\Big] \sqrt{E^2 - D^2 - P_1^2 - P_2^2} \ . \nonumber \\
\end{eqnarray}
 
Now these conserved charges hold the relationship 
\begin{eqnarray}
    (J_1)_{reg} &=& \sqrt{(K_1)_{reg}^2 - \frac{3\lambda}{\pi^2} \sin^2 \Big( \frac{(\Delta\phi_1)_{reg}}{2}\Big)} \ , \nonumber \\ (J_2)_{reg} &=& \sqrt{(K_2)_{reg}^2 - \frac{3\lambda}{\pi^2} \sin^2 \Big( \frac{(\Delta\phi_2)_{reg}}{2}\Big)} \ .
\end{eqnarray}
Like previous subsection here also we obtain two sets of single spike relations arising from two orthogonal spheres of I-brane and each single spike is of the form (\ref{ss1}) instead of (\ref{ss2}). 

\section{Summary and conclusion}\label{sec4}
In this short paper, we revisit the rigidly rotating probe strings in I-brane background with Polyakov action for the string. Our aim was to study the rotating string solution over the whole space of I-brane without any restriction. Unlike Nambu-Goto action, Polyakov action for the string allows us decouple the two orthogonal spheres that arise from the near horizon limit of orthogonal stacks of NS5-branes. This might be because the Polyakov action for the string is more symmetric than the Nambu-Goto action for the string and this extra symmetry allows us to decouple the spheres. This allows us to study the rotating strings in both the spheres simultaneously, however the motion is visibly complicated in nature. In this case we consider three different cases in the parameter space of integration constants and obtain the giant magnon like dispersion relation in one case and single spike in other cases. This observation leads us to affirm that single spike or spiky string is more generalized solutions of rotating strings while giant magnon is a special case of spiky string solutions. 
\medskip

We also obtain two sets of giant magnon and single spike relations arising from the orthogonal spheres, we have seen this type of behaviour while studying rotating string in curved NS5-branes in \cite{Papadopoulos:1999tw, Biswas:2013ela}. This type of behaviour was not seen while studying rotating string using Nambu-Goto action under the restriction $\theta_1=\theta_2=\theta$. This might be because under this restriction we were studying rigidly rotating string on an effective sphere of the form $d\Omega^2 = 2d\theta^2 + \sin^2\theta (d\phi_1^2 + d\phi_2^2) + \cos^2\theta(d\psi_1^2 + d\psi_2^2)$. With the identification $d\phi_1^2+d\phi_2^2=2d\phi^2$ and $d\psi_1^2+d\psi_2^2=2d\psi^2$ and rescaling $d\Omega^2 \to d\Omega^2/2$ we effectively studied rotating strings on $\mathbb{R}_t\times S^3$ and hence we obtain one giant magnon or single spike relation. Whereas Polyakov action for the string didn't require such restriction we could study the rotating strings on $\mathbb{R}_t\times S^3\times S^3$ and hence we obtain two sets of giant magnon and single spike relations for two orthogonal spheres.
\medskip

We have mentioned that a similar kind of difficulty was faced while studying rotating D1-strings on two orthogonal stacks of D5-branes \cite{Banerjee:2016avv}, where we determined the string equations of motion using DBI-action. This scenario is explicitly S-dual to the present case, i.e. F1 strings in stacks of NS5-branes. Like Nambu-Goto action DBI action is of the square root form and this action does not bear enough symmetries to decouple the spheres and we had to study the rotating D1-string under the same restriction of $\theta_1=\theta_2=\theta$. This restriction can be avoided if one uses an action for D1-string which is of the form of Polyakov action. As the dual field theory is not yet understood till now, it is quite difficult to comment on the dual operators from these dispersion relations. An interesting arena that we could explore in this regard could be the oscillating circular string, also known as ``Pulsating Strings" \cite{Minahan:2002rc} in the literature. We hope to focus on these investigations in the near future. 

\section*{Acknowledgements} 
It's a pleasure to thank Aritra Banerjee for some useful discussions which help to gain deeper insights and improves the manuscript.

\end{document}